\documentclass[12pt]{article}

\usepackage{graphicx}
\usepackage{cite}

\title{Statistical physics in
deformed spaces with minimal length}
\author{T.V. Fityo\footnote{E-mail: fityo@ktf.franko.lviv.ua}\\
{\small Chair of Theoretical Physics, Ivan Franko National University of Lviv,}\\
{\small 12 Drahomanov St., Lviv, UA-79005, Ukraine} }

\newcommand{\pd}[2]{\frac{\partial #1}{\partial #2}}
\newcommand{\jac}[2]{\frac{\partial\left(#1\right)}{\partial\left(#2\right)}}

\begin{document}


\maketitle

\abstract{We considered the thermodynamics in spaces with deformed
commutation relations leading to the existence of minimal length.
We developed a classical method of the partition function
evaluation. We calculated the partition function and heat capacity
for ideal gas and harmonic oscillators using this method. The
obtained results are in good agreement with the exact quantum
ones. We also showed that the minimal length introduction reduces
degrees of freedom of an arbitrary system in the high temperature
limit significantly.}



\section{Introduction}

Now it is widely accepted that quantum gravity and string theory
lead to minimal length, i.e. there must exist a lower bound  to
the possible resolution of the distance \cite{Mead64, Gross88,
Amati89, Maggiore93, Garay95, Witten96, Scardigli99, Szabo03}. The
minimal length appears due to a modification of the standard
Heisenberg uncertainty relation:
\begin{equation}
\Delta X\Delta P\ge\frac\hbar2(1+\beta \Delta P^2+\dots),
\end{equation}
where $\beta$ is a positive parameter. This modified relation
implies that uncertainty of coordinate $\Delta X$ is always larger
than $\Delta X_{min}=\hbar\sqrt\beta$. It was shown that such a
modified uncertainty relation can be obtained as a natural
consequence of a specific modification of the usual commutation
relation \cite{Maggiore93b, Maggiore94, Kempf94, Kempf95,
Kempf97b}. We would like to note that modification of the
commutation relation is only one of the possible ways of
introducing minimal length. For a review of different approaches
to theories with the minimal length see \cite{Hossenfelder06}.

The deformed commutation relation according to Kempf et~al.
\cite{Kempf94, Kempf95, Kempf97b} may read in one-dimensional
space
\begin{equation}
[X,P]=i\hbar(1+\beta P^2)
\end{equation}
or in $D$-dimensional space ($i,j\in {1\dots D}$)
\begin{equation}\label{ds1}
[X_i,P_j]=i\hbar((1+\beta P^2)\delta_{ij}+\beta'P_iP_j),\qquad
[P_i,P_j]=0,
\end{equation}
\begin{equation}\label{ds3}
[X_i,X_j]=i\hbar\frac{2\beta-\beta'+(2\beta+\beta')\beta
P^2}{1+\beta P^2}\left(P_iX_j-P_jX_i \right).
\end{equation}

It is believed widely that the minimal length can be helpful in
the context of divergencies regularization in different areas of
theoretical physics \cite{Garay95}. Contrary to the previous
theories, where the minimal length appears phenomenologically, in
our case the minimal length appears as a natural consequence of
modified commutation relations. An explicit representation of
position and momentum operators is known and the existence of
possible regularization can be verified from the first principles.
It was shown that deformed commutation relations really might lead
to regularization in quantum field theory \cite{Kempf97}.
Regularization with the help of deformed commutation relations for
specific eigenvalue problems was studied in \cite{Fityo06b,
Bouaziz07}.

Other implications of non-zero minimal length were considered in
the context of the following problems: harmonic oscillator
\cite{Kempf95, Kempf97b, Chang02, Quesne03, Quesne04, Quesne05},
hydrogen atom \cite{Brau99, Benczik05, Stetsko06, Stetsko06b},
gravitational quantum well \cite{Brau06}, the Casimir effect
\cite{Nouicer05, Harbach06}, particles scattering
\cite{Hossenfelder06b, Stetsko07}. In several of the mentioned
papers attempts to estimate the value of minimal length upper
bound were made by comparing theoretical predictions and
experimental data. In \cite{Brau99, Benczik05, Stetsko06,
Stetsko06b} considering the Lamb shift the authors estimated
$\Delta X_{min}\le 10^{-16}\dots 10^{-17}$m, analysis of electron
motion in a Penning trap also gives $\Delta X_{min}\le 10^{-16}$m
\cite{Chang02}, consideration of neutron motion in the
gravitational field \cite{Brau06} gives large $\Delta X_{min}\le
2.4 \cdot 10^{-9}$m due to significant experimental errors. In our
analysis we will use $\Delta X_{min} = 10^{-16}$m.

Consideration of some theoretical physics topics requires the use
of statistical methods. The purpose of this paper is to develop
statistical physics with some general form of deformed commutation
relation. There are several papers on the subject. In
\cite{Rama01} the statistical physics is constructed by
modification of elementary cell of space volume according to
modification of the commutation relation. In that paper a very
strong restriction $[X_i,X_j]=0$ was assumed. A similar analysis
was performed in \cite{Lubo03,Lubo03b}. Density of states and
black-body radiation were considered in \cite{Chang02b} with the
help of Liouville theorem analog. The microcanonical approach to
ideal gas was considered in \cite{Nozari06}.

In this paper we develop a method for the consideration of
thermodynamical properties of the system with arbitrary
commutation relation between momenta and positions operators. We
approximate the value of the partition function with the help of a
semiclassical approach. We may expect that such an approximation
is applicable since the WKB approximation gives correct
predictions for one-dimensional eigenvalues problems with deformed
commutation relations \cite{Fityo06, Fityo07}. We apply this
method to the particular case, namely to the three-dimensional
Kempf's commutation relation (\ref{ds1}--\ref{ds3}) and consider
implications of the minimal length on thermodynamics.

\section{A method}

We consider a system consisting of $N$ identical non-interacting
particles in the external field $U$. The behavior of each particle
may be described by Schr\"odinger Hamiltonian
\begin{equation}\label{dsH}
H=\frac{P^2}{2m}+U(\mathbf{X}).
\end{equation}
We use this Hamiltonian to analyze examples, but the developed
method can be applied to a system described by Hamiltonian of any
form.

In this paper we assume that the partition function of a quantum
system can be evaluated as
\begin{equation}\label{dsPFdd}
Z=\sum_n e^{-E_n/T},
\end{equation}
where $E_n$ are eigenvalues of the Hamiltonian, $n$ denotes
different states. This formula has the same form either in the
deformed or in the non-deformed cases. In the non-deformed case
one can use a semiclassical approximation for the partition
function
\begin{equation}\label{ds4}
Z=\int e^{-H/T}(dx)(dp),
\end{equation}
where $(dx)$ means $dx_1 dx_2\dots dx_D$. Since it is the
non-deformed case, the coordinate and the momentum variables $x_i$
and $p_j$ are canonically conjugated $\{x_i,p_j\}=\delta_{ij}$,
$\{x_i,x_j\}=\{p_i,p_j\}=0$. Here and below small variables $x$,
$p$ denote canonically conjugated variables.

We show below that the existence of deformation becomes
perceptible only for very high temperature. For high temperature
the difference between Bose-Einstein, Fermi-Dirac or Boltzmann
statistics is irrelevant, therefore we use Boltzmann one as the
simplest. One of the aims of this paper is to generalize formula
(\ref{ds4}), which gives correct predictions for high temperature,
to the case of the deformed commutation relation. Peculiarities of
Bose-Einstein and Fermi-Dirac statistics for the particular form
of deformed commutation relations were considered in \cite{Lubo03,
Lubo03b}.

Let us consider the general case of deformed commutation relation
\begin{equation}\label{ds1g}
[X_i,P_j]=i\hbar f_{ij}(X,P),
\end{equation}
\begin{equation}\label{ds2g}
[P_i,P_j]=i\hbar h_{ij}(X,P),
\end{equation}
\begin{equation}\label{ds3g}
[X_i,X_j]=i\hbar g_{ij}(X,P).
\end{equation}
In the above expressions operators $X_i$, $P_j$ correspond to the
same particle. Operators describing different particles commute.
The deformation functions $f_{ij}$, $g_{ij}$, $h_{ij}$ are
restricted according to the properties of commutators:
bilinearity, the Leibniz rules and the Jacobi identity. The
investigation on this subject for some special cases of the
deformation function can be found in \cite{Kempf97b, Kempf97}.

In the classical limit $\hbar\to0$ the deformed commutation
relations (\ref{ds1g}-\ref{ds3g}) lead to the deformed Poisson
bracket \cite{Chang02b,Benczik02,Frydryzhak03}
\begin{eqnarray}\label{s2m}
\{X_i,P_j\}=f_{ij}(X,P), \quad \{P_i,P_j\}= h_{ij}(X,P), \quad
\{X_i,X_j\}=g_{ij}(X,P).
\end{eqnarray}
These Poisson brackets possess the same properties as the quantum
mechanical commutators (\ref{ds1g}--\ref{ds3g}), namely, they are
anti-symmetric, bilinear, and satisfy the Leibniz rules and the
Jacobi identity \cite{Benczik02,Frydryzhak03}. According to
Darboux theorem \cite{Arnold89} it is always possible to choose
such auxiliary canonically conjugated variables $x_i$ and $p_i$
that $X_i$ and $P_i$ as functions of $x_i$ and $p_i$ satisfy
equations (\ref{s2m}). Where, just as an example,
\begin{eqnarray}\label{s3}
\{X_i,P_j\}=\sum_{k=1}^D\pd{X_i}{x_k}\pd {P_j}{p_k}-\pd
{P_j}{x_k}\pd {X_i}{p_k}= f_{ij}(X,P).
\end{eqnarray}

Then, the initial Hamiltonian (\ref{dsH}) can be considered as the
function of $(x,p)$ and the one particle partition function can be
evaluated according to formula (\ref{ds4}). In appendix
\ref{appJac} we show that the Jacobian $J=\jac{X,P}{x,p}$ can
always be expressed as a combination of Poisson brackets
(\ref{s2m}). It is an important result since it gives a
possibility to calculate the partition function without
introducing canonically conjugated auxiliary variables. Namely,
\begin{equation}\label{dspf}
Z=\int e^{-H(X,P)/T}\frac{(dX)(dP)}J.
\end{equation}
This equation is an analog of formula (\ref{ds4}) for the deformed
case. It is an essential and expected result that the partition
function does not depend on auxiliary variables since $J$ depends
only on the structure of the Poisson brackets (\ref{s2m}) which,
together with the Hamiltonian, define all the properties of the
system.

Expression (\ref{dspf}) can be considered as a semiclassical
approximation of partition function (\ref{dsPFdd}). It is well
known that the semiclassical approximation is good for large
quantum numbers (it corresponds to high temperature regime). We
will observe this situation with the help of the harmonic
oscillators ensemble example: heat capacities calculated according
to formulae (\ref{dsPFdd}) and (\ref{dspf}) coincide for high $T$.

\section{Examples}

In this section we analyze the effect of deformation
(\ref{ds1}--\ref{ds3}) on the thermodynamical quantities. Since
Jacobian (\ref{ds1a}) does not depend on coordinates and
Hamiltonian has the form (\ref{dsH}) evaluation of the
one-particle partition function (\ref{dspf}) can be separated in
two parts: integrating over $(X)$ and over $(P)$. Namely,
\begin{equation}\label{ss1}
Z=\int(dX)\exp\left[-\frac{U(\mathbf{X})}{T}\right] \int(dP)
\frac{ \exp\left[-\frac{P^2}{2mT}\right]} {(1+\beta
P^2)^2(1+(\beta+\beta')P^2)}.
\end{equation}
Note, that during the transition from quantum expression for the
partition function (\ref{dsPFdd}) to the classical one
(\ref{ss1}), we implicitly assume that $\beta$ and $\beta'$ are
independent of $\hbar$, i.e. we keep the deformation parameters
fixed as $\hbar\to0$. Some comments on this topic can be found in
\cite{Benczik02}.

It is easy to see that deformation changes the partition function
perceptibly if $\beta m T\sim 1$, $\beta' m T\sim 1$. For electron
gas with the minimal length $10^{-16}$m $\beta m T\sim 1$ if
$T\sim10^{18}$K. The very high value of this temperature justifies
our choice of Boltzmann statistics.

For low temperatures $\beta m T\ll 1$, $\beta' m T\ll 1$ the
partition function of one particle (\ref{ss1}) simplifies to
\begin{equation}
Z=Z_0(1-3(3\beta+\beta')mT+o(T)),
\end{equation}
where $\displaystyle Z_0=\int(dX)\exp\left[-U/T\right]
\int(dP)\exp\left[-P^2/2mT\right]$ is the partition function of
one particle described by the same Hamiltonian for the
non-deformed case. The existence of deformation leads to a change
of internal energy and heat capacity in such a way
\begin{eqnarray}
E=E_0-3(3\beta+\beta')NmT^2+o(T^2),\\
C=C_0-6(3\beta+\beta')NmT +o(T).
\end{eqnarray}

For large $T$: $\beta m T\gg 1$ the partition function of one
particle, internal energy and heat capacity are
\begin{eqnarray}
Z=Z_0\frac{\pi^2(2\pi mT)^{-3/2}}
{\sqrt\beta(\sqrt\beta+\sqrt{\beta+\beta'})^2}\left(1+O\left(\frac1T\right)\right),\\
E=E_0-\frac32NT+O(1),\qquad C=C_0-\frac32N+O\left(\frac1T\right).
\end{eqnarray}
This result will be verified on the quantum level for two examples
(ideal gas and ensemble of 3D harmonic oscillators). We see that
the presence of deformation freezes three degrees of freedom at
high temperature. In section \ref{appHigh} we show that such a
freezing is a natural consequence of the minimal length
introduction.

\subsection{Ideal gas} Hamiltonian of one particle is $H=P^2$.
We consider that the entire system is confined in volume $V$, i.e.
for each particle $\displaystyle \int (dX)=V$.

For low temperatures, the one-particle partition function reads
\begin{equation}
Z=V(2\pi mT)^{3/2}(1-3(3\beta+\beta')mT+o(T)).
\end{equation}
It leads to the following expressions for internal energy and heat
capacity
\begin{eqnarray}
E=\frac32NT-3(3\beta+\beta')NmT^2+o(T^2),\quad
C=\frac32N-6(3\beta+\beta')NmT+o(T).
\end{eqnarray}

In limit $T\to\infty$ the one-particle partition function is
\begin{equation}
Z=\frac{V\pi^2}
{\sqrt\beta(\sqrt\beta+\sqrt{\beta+\beta'})^2}\left(1-\frac1{2mT\beta}
\frac{2\sqrt\beta+\sqrt{\beta+\beta'}}{\sqrt{\beta+\beta'}}
+o\left(\frac1T\right)\right).
\end{equation}
Here we take into account the second term of the series expansion
to obtain the second correction to the internal energy expression
(the leading term, which must be proportional to $T$, is zero):
\begin{equation}\label{de8}
E=\frac1{2m\beta}\frac{2\sqrt\beta+ \sqrt{\beta+\beta'}} {
\sqrt{\beta+\beta'}}+o(1).
\end{equation}
The last equation can be explained with the help of the quantum
analysis. Although the problem of the particle in the box in the
deformed case has not been solved exactly yet, the preliminary
analysis shows that there exists only a finite amount of bound
states \cite{Brout99,Detournay02}. For high temperature it must
lead to the situation that all particles occupy the highest energy
level and thus, heat capacity must tend to $0$ as we see from
(\ref{de8}).

It is interesting to note that equation of state is the same as in
the non-deformed case and its form does not depend on the
temperature value. Namely, it reads
\begin{equation}
pV=NT.
\end{equation}

\subsection{3D harmonic oscillators}\label{hoex}
One-particle harmonic oscillator Hamiltonian is, as usual,
$\displaystyle H=\frac{P^2}{2m}+\frac{m\omega^2}2X^2$. For low
temperature expression (\ref{ss1}) gives the following formula for
the partition function
\begin{equation}\label{de10}
Z=\left(\frac{2\pi T}{\omega}\right)^3(1-3(3\beta+\beta')mT+o(T))
\end{equation}

The eigenvalue problem of the harmonic oscillator has been solved
exactly in \cite{Chang02} for the first time and the spectrum
reads
\begin{eqnarray}\label{de11}
E_{nl}=\hbar\omega\left( \left(n+\frac32\right) \sqrt{1+
m^2\omega^2\hbar^2 \left[\beta^2l(l+1)+
\frac{(3\beta+\beta')^2}4\right]} \right.\nonumber\\
\left. +\frac{m\omega\hbar}2\left[ (\beta+\beta')
\left(n+\frac32\right)^2 + (\beta-\beta')
\left(l(l+1)+\frac94\right) +\frac32\beta' \right] \right).
\end{eqnarray}
We calculate the one-particle partition function using formula
(\ref{dsPFdd}) in linear approximation over $\beta$, $\beta'$
exactly. This expression is rather cumbersome, but in limit
$\hbar\to0$ we obtain simple expression for it
\begin{equation}
Z=\frac{T^3}{\hbar^3\omega^3}(1-3(3\beta+\beta')mT)+O(1/\hbar^2).
\end{equation}
The leading term of the last expression differs from (\ref{de10})
only by constant factor $(2\pi\hbar)^3$ which does not have any
effect on thermodynamics and which we omit in the initial
definitions of approximation (\ref{ds4}, \ref{dspf}).

For large $T$ expression (\ref{ss1}) gives
\begin{equation}
Z=\left(\frac{2\pi T}{m \omega^2}\right)^{3/2}
\frac{\pi^2}{\sqrt\beta(\sqrt\beta+\sqrt{\beta+\beta'})^2}
\left(1+O\left(\frac1T\right)\right),
\end{equation}
\begin{equation}\label{de13}
E=\frac32NT+O(1).
\end{equation}
This result can be easily explained and verified as follows. The
energies with large $n$ give the main contribution to the
partition function at high temperatures. In the deformed case for
large $n$ according to expression (\ref{de11}) $\displaystyle
E_{nl}\sim \frac{m\omega\hbar}2(\beta+\beta')n^2$. Such an energy
dependence on the main quantum number is the same as for a
particle in the box for the non-deformed case. Thus, it is an
expected result to reproduce particles-in-the-box internal energy
dependence on the temperature (\ref{de13}).

\begin{figure}[htb!]
\centerline{\includegraphics[width=0.6\textwidth,clip,angle=-0]{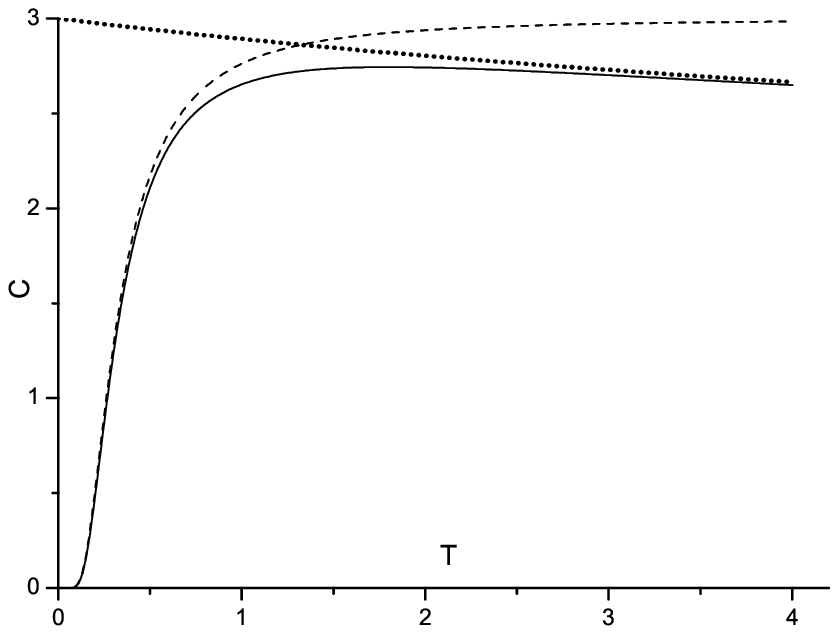}}
\centerline{\includegraphics[width=0.6\textwidth,clip,angle=-0]{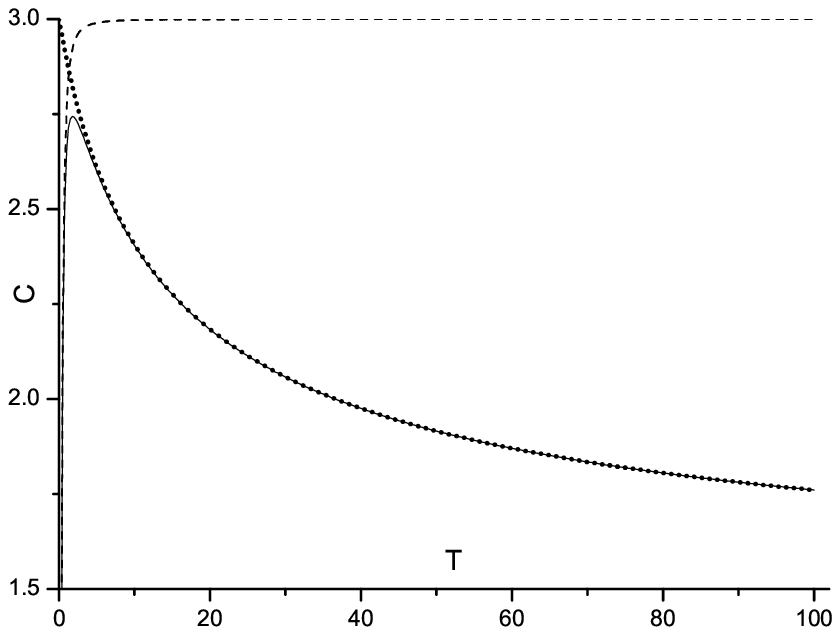}}
\caption{Temperature dependence of heat capacity of the harmonic
oscillators ensemble per one particle. $\beta=\beta'=0.01$,
$\hbar=\omega=2m=1$, $T$ is given in $\hbar\omega$ units. Dashed
line --- heat capacity in the non-deformed case
($\beta=\beta'=0$), solid line --- heat capacity calculated
according to the direct definition of the partition function
(\protect\ref{dsPFdd}), dotted line --- according to classical
approximation (\protect\ref{dspf}). }\label{defStat.eps}
\end{figure}

Heat capacity dependence on temperature is plotted on
Fig.~\ref{defStat.eps}. For high $T$ heat capacity tends to $3/2$
what follows from (\ref{de13}). One can see that approximation
(\ref{dspf}) gives predictions very closely to the exact result
for high $T$. For selected values of parameters
$\beta=\beta'=0.01$, $\hbar=\omega=2m=1$ it was expected that
effects implied by the deformed commutation relation become
significant for $\displaystyle T=\frac1{\beta m}\sim 100$. As one
can see this implication appears to be important for a much lower
temperature, namely for $T=1\div2$.

\section{Heat capacity at high temperature and the minimal
length}\label{appHigh}

For a wide class of physically important systems, Hamiltonian can
be expressed as a sum of kinetic and potential energy. In this
section we estimate the contribution of kinetic energy to heat
capacity in the limit $T\to\infty$.

Let us consider the non-deformed case. In the case of high
temperature only large values of momenta contribute to the
partition function. For large $p$ the Hamiltonian can be
approximated as $H = \alpha p^n$ (for Schr\"odinger Hamiltonian
$\alpha=\frac1{2m}$, $n=2$; for Dirac Hamiltonian $\alpha=c$,
$n=1$). So,
\begin{equation}
Z(T\to\infty)\sim \int^{\infty}e^{-\frac{\alpha p^n}T}p^{D-1}dp=
T^{\frac Dn}\alpha^{-\frac Dn}\int^{\infty}e^{-x}x^{\frac{D-1}n}
dx
\end{equation}
Such a dependence on $T$ means that $\displaystyle
C(T\to\infty)=\frac Dn$, which reproduces several well-known
results.

Let us consider deformed commutation relations leading to the
minimal length. In a one-dimensional case
$$[X,P]=i\hbar(1+f(X,P))$$
and Schr\"odinger uncertainty relation reads
\[
\Delta X\ge\frac\hbar2\left(\frac1{\Delta P}+ \frac{\langle
f(X,P)\rangle}{\Delta P} \right)\ge \frac\hbar2\left(\frac1{P}+
\frac{\langle f(X,P)\rangle}P\right),
\]
where $P=\sqrt{\langle P^2\rangle}\ge \Delta P$. Therefore, a
nonzero minimal length exists if $\displaystyle
\lim_{P\to\infty}f(X,P)$ grows as $P$ or faster for large momentum
values. Then the Jacobian $J=1+f$ for large $P$ must grow as $P$
or faster. In the $D$-dimensional case it must grow as $P^D$ or
faster.

Let us first consider the case $J\sim P^D$ and denote by $P_0$
such an intermediate value of momentum that $H\approx\alpha P^n$,
$J\approx\gamma P^D$ for $P\ge P_0$, but $H\ll T$ for $P\le P_0$.
Then the partition function reads
\begin{equation}
Z=\int\limits_0^\infty e^{-H/T}\frac{P^{D-1}dP}{J}\approx
\int\limits_0^{P_0}\frac{P^{D-1}dP}{J}+\frac1\gamma
\int\limits_{P_0}^\infty e^{-\alpha
P^n/T}\frac{dP}P=\textrm{const}+\frac1\gamma\ln\frac T{\alpha
P_0^n}
\end{equation}
Heat capacity can be easily calculated and $C(T\to\infty)=0$.

A similar procedure can be performed when the Jacobian grows
faster than $P^D$. It gives $Z(T\to\infty)=\textrm{const}$ and
consequently $C(T\to\infty)=0$. So, in any case leading to the
minimal length $C(T\to\infty)=0$.

We may conclude that for a wide class of Hamiltonians being
proportional to a power of momentum for large momentum values and
deformed commutation relations leading to the nonzero position
uncertainty in the limit $T\to\infty$ the partition function of
the system is as $\ln T$ (the Jacobian grows as $P^D$) or does not
depend on temperature (the Jacobian grows faster). Case of the
constant partition function was discussed in the context of the
string theory from the duality point of view \cite{Atick88}. A
similar effect of degrees of freedom reduction was observed for
two model Hamiltonians with particular deformed commutation
relations in the context of grand canonical approach
\cite{Rama01}.

\section{Conclusions}

In this paper we developed a general method for the consideration
of thermodynamical properties of an arbitrary system of
non-interacting particles with deformed commutation relations. To
calculate a classical expression for the partition function we
express the Jacobian of making a change from arbitrary canonically
conjugated variables to the initial variables as the combination
of deformed Poisson brackets.

We calculate the partition function for two model systems, namely
ideal gas and ensemble of harmonic oscillators and considered
corresponding thermodynamical properties. We showed that such a
derived partition function is in qualitative agreement with the
exact quantum expression in the limit of high temperature. Quite
interesting consequence of the minimal length introduction is
decreasing of the heat capacities of these model systems to $0$
and $3/2$ for high temperatures. We managed to show that such a
reduction of the heat capacity occurs (when compared with the
corresponding non-deformed system) for each system with arbitrary
deformed commutation relations leading to the minimal length.
Namely, the minimal length introduction completely removes
translation degrees of freedom for very high temperature.

\section{Acknowledgments}
The author is grateful to Prof.~V.~M.~Tkachuk and
Dr.~A.~A.~Rovenchak for numerous helpful discussions and accurate
reading the ma\-nus\-cript.

\appendix

\section{Expressing Jacobian as a combination of the Poisson
brackets}\label{appJac}

Let us denote $X_i=A_{2i-1}$, $P_i=A_{2i}$, $A_j$ derivative with
respect to $x_i$ we denote $A_{j,2i-1}$, with respect to $p_i$ as
$A_{j,2i}$. Then
$$\left\{A_i,A_j\right\}=\sum_{k=1}^D (A_{i,2k-1} A_{j,2k}- A_{i,2k} A_{j,2k-1}). $$
Let us prove the following identity
\begin{equation}\label{mnIden}
J=\jac{X_1,P_1,\dots,X_D,P_D}{x_1,p_1,\dots,x_D,p_D} =
\frac1{2^DD!}\sum_{i_1,\dots i_{2D}=1}^{2D} \varepsilon_{i_1\dots
i_{2D}}\{A_{i_1},A_{i_2}\} \dots \{A_{i_{2D-1}},A_{i_{2D}}\},
\end{equation}
where $\varepsilon_{i_1\dots i_{2D}}$ is the Levi-Civita symbol.
The right-hand of this identity is equal
\begin{eqnarray}\label{scIden}
&\displaystyle\frac1{2^DD!}\sum_{i_1,\dots i_{2D}=1}^{2D}
\varepsilon_{i_1\dots i_{2D}}\{A_{i_1},A_{i_2}\} \cdot\dots\cdot
\{A_{i_{2D-1}},A_{i_{2D}}\} = \frac1{2^DD!}\sum_{i_1,\dots
i_{2D}=1}^{2D} \varepsilon_{i_1\dots i_{2D}}\nonumber\\
&\displaystyle\sum_{j_1=1}^D(A_{i_1,2j_1-1}A_{i_2,2j_1}-A_{i_1,2j_1}A_{i_2,2j_1-1})
\dots \sum_{j_D=1}^D(A_{i_{2D-1},2j_D-1}A_{i_{2D},2j_D}-
A_{i_{2D-1},2j_D}A_{i_{2D},2j_D-1})\nonumber\\
&\displaystyle= \frac1{D!}\sum_{j_1,\dots j_D} \sum_{i_1,\dots
i_{2D}} \varepsilon_{i_1\dots i_{2D}} A_{i_1,2j_1-1}A_{i_2,2j_1}
\dots A_{i_{2D-1},2j_D-1}A_{i_{2D},2j_D}.
\end{eqnarray}
In the latter equality we take into account that the Levi-Civita
symbol is antisymmetric with respect to any indexes permutation,
thus the following property holds $\displaystyle
\sum_{i_1,i_2}\varepsilon_{i_1\dots i_{2D}} A_{i_1,2j_1-1}
A_{i_2,2j_1} = -\sum_{i_1,i_2}\varepsilon_{i_1\dots i_{2D}}
A_{i_1,2j_1} A_{i_2,2j_1-1}$.

From the fact that
\begin{eqnarray}
\sum_{i_1,\dots i_{2D}} \varepsilon_{i_1\dots i_{2D}}
A_{i_1,2j_1-1}A_{i_2,2j_1} \dots
A_{i_{2D-1},2j_D-1}A_{i_{2D},2j_D}=\nonumber\\
\det \left(
\begin{array}{ccccc}
  A_{1,2j_1-1} & A_{1,2j_1} & \dots & A_{1,2j_D-1} & A_{1,2j_D}\\
  A_{2,2j_1-1} & A_{2,2j_1} & \dots & A_{2,2j_D-1} & A_{2,2j_D}\\
  \vdots & \vdots & \ddots & \vdots & \vdots\\
   A_{2D,2j_1-1} & A_{2D,2j_1} & \dots & A_{2D,2j_D-1} & A_{2D,2j_D}
\end{array}
\right)\nonumber
\end{eqnarray}
and using the determinant properties it becomes obvious that in
the last formula of equation (\ref{scIden}) only terms with
different $j$ give the contribution to the final result. All the
terms with $j_1\ne j_2\ne \dots\ne j_{2D}$ are equal. The total
number of these terms equals $D!$. Thus,
\begin{eqnarray}
\frac1{D!}\sum_{j_1,\dots j_D} \sum_{i_1,\dots i_{2D}}
\varepsilon_{i_1\dots i_{2D}} A_{i_1,2j_1-1}A_{i_2,2j_1} \dots
A_{i_{2D-1},2j_D-1}A_{i_{2D},2j_D}=\nonumber\\\sum_{i_1,\dots
i_{2D}} \varepsilon_{i_1\dots i_{2D}} A_{i_1,1}A_{i_2,2} \dots
A_{i_{2D-1},2D-1}A_{i_{2D},2D}=\det(A_{ij}),
\end{eqnarray}
which proves identity (\ref{mnIden}).

The right-hand side of expression (\ref{mnIden}) contains $(2D)!$
terms, each of them is a product of $D$ Poisson brackets. Due to
the skew-symmetry of Poisson bracket some of the terms are equal
and the total amount of terms can be reduced to $(2D-1)!!$ terms.
Below we enlist Jacobians for one-, two- and three-dimensional
cases.

In a one-dimensional case the following expression
\begin{equation}
\jac{X,P}{x,p}=\left\{X,P\right\}
\end{equation}
is obvious.

In a two-dimensional case
\begin{equation}
\jac{X_1,P_1,X_2,P_2}{x_1,p_1,x_2,p_2}=\left\{X_1,P_1\right\}\left\{X_2,P_2\right\}-
\left\{X_1,X_2\right\}\left\{P_1,P_2\right\}-\left\{X_1,P_2\right\}\left\{X_2,P_1\right\}
\end{equation}
can be checked by hand.

A three-dimensional case:
\begin{eqnarray}\label{3DKempfJ}
\jac{X_1,P_1,X_2,P_2,X_3,P_3}{x_1,p_1,x_2,p_2,x_3,p_3}=
\left\{X_1,P_1\right\}\left\{X_2,P_2\right\}\left\{X_3,P_3\right\}-\nonumber\\
\left\{X_1,P_3\right\}\left\{P_1,P_2\right\}\left\{X_2,X_3\right\}-
\left\{X_1,P_2\right\}\left\{X_2,P_1\right\}\left\{X_3,P_3\right\}-\nonumber\\
\left\{X_1,P_3\right\}\left\{X_2,P_2\right\}\left\{X_3,P_1\right\}-
\left\{X_1,P_1\right\}\left\{X_2,P_3\right\}\left\{X_3,P_2\right\}+\nonumber\\
\left\{X_1,X_2\right\}\left\{P_1,P_3\right\}\left\{X_3,P_2\right\}+
\left\{X_1,P_3\right\}\left\{X_2,P_1\right\}\left\{X_3,P_2\right\}-\nonumber\\
\left\{X_1,X_2\right\}\left\{P_2,P_3\right\}\left\{X_3,P_1\right\}+
\left\{X_1,P_2\right\}\left\{X_2,X_3\right\}\left\{P_1,P_3\right\}-\nonumber\\
\left\{X_1,X_3\right\}\left\{P_1,P_3\right\}\left\{X_2,P_2\right\}+
\left\{X_1,X_3\right\}\left\{X_2,P_1\right\}\left\{P_2,P_3\right\}+\nonumber\\
\left\{X_1,X_3\right\}\left\{P_1,P_2\right\}\left\{X_2,P_3\right\}-
\left\{X_1,X_2\right\}\left\{P_1,P_2\right\}\left\{X_3,P_3\right\}-\nonumber\\
\left\{X_1,P_1\right\}\left\{X_2,X_3\right\}\left\{P_2,P_3\right\}+
\left\{X_1,P_2\right\}\left\{X_2,P_3\right\}\left\{X_3,P_1\right\}.
\end{eqnarray}
This expression can be checked with the help of the computer.

It is easy to see that to obtain such a formula one needs to start
from the term $\left\{X_1,P_1\right\} \cdot\dots\cdot
\left\{X_D,P_D\right\}$ and add to it all possible permutations.
Factor multiplying each term is either $+1$ for even permutation
and $-1$ for odd permutation. In the above expression all
variables are ordered, i.e. in each Poisson bracket $X_i$ is
before $P_j$, $X_i$ is before $X_j$ if $j>i$, and $P_i$ is before
$P_j$ if $j>i$.

For deformation (\ref{ds1}--\ref{ds3}) Jacobian (\ref{3DKempfJ})
simplifies to
\begin{equation}\label{ds1a}
\jac{X_1,P_1,X_2,P_2,X_3,P_3}{x_1,p_1,x_2,p_2,x_3,p_3}=(1+\beta
P^2)^2(1+(\beta+\beta')P^2).
\end{equation}

In the case of small deviation of the deformed Poisson brackets
(\ref{s2m}) from the canonical ones ($f_{ij}-\delta_{ij},\
h_{ij},\ g_{ij}\approx 0$) expression (\ref{mnIden}) can be
simplified significantly in the linear approximation over these
deviations. It is easy to see that in this case only the first
term $\left\{X_1,P_1\right\} \cdot\dots\cdot
\left\{X_D,P_D\right\}$ contributes to the Jacobian and
$$
J=\prod_{i=1}^Df_{ii}(X,P)=1+\sum_{i=1}^D(f_{ii}(X,P)-1).
$$
Such an expression is useful for the analysis of low temperature
behavior.

\end{document}